\documentclass[useAMS,usenatbib]{mn2e}
\usepackage{epsfig}
\newcommand\msun{\rm{M_{\odot}}}
\title[Hydrodynamical simulations of Galactic fountains I]{Hydrodynamical 
simulations of Galactic fountains I: evolution of single fountains.}
\author[C. Melioli, F. Brighenti, A. D'Ercole, E.M. de Gouveia Dal Pino]
{C. Melioli,$^{1}$\thanks{E-mail:
cmelioli@astro.iag.usp.br; fabrizio.brighenti@unibo.it; annibale.dercole@oabo.inaf.it;
dalpino@astro.iag.usp.br} 
F. Brighenti$^{1}$, A.D'Ercole$^{2}$ and E.M. de Gouveia Dal Pino$^{3}$\\
$^{1}$Dipartimento di Astronomia, Universit\`a di Bologna, via Ranzani 1, 40126 Bologna, Italy\\
$^{2}$INAF-Osservatorio Astronomico di Bologna, via Ranzani 1, 40126 Bologna, Italy\\
$^{3}$IAG, Universidade de S\~ao Paulo, Rua do Mat\~ao 1226, 05508-090 S\~ao Paulo, Brazil}
\begin{document}

\date{Accepted ... Received ...; in original form ...}

\pagerange{\pageref{firstpage}--\pageref{lastpage}} \pubyear{2008}

\maketitle

\label{firstpage}

\begin{abstract}

The ejection of the gas out of the disk in late-type galaxies
is related to star formation 
and is due mainly to Type II supernovae.
In this paper we studied in detail the development of the Galactic
fountains in order to understand their dynamical evolution and their
influence in the redistribution of the freshly delivered metals over
the disk.  To this aim, we performed a number of 3D hydrodynamical
radiative cooling simulations of the gas in the Milky Way where the whole
Galaxy structure, the Galactic differential rotation and the
supernovae explosions generated by a single OB association are
considered.  A typical fountain powered by 100 Type II supernovae may
eject material up to $\sim 2$ kpc which than collapses back mostly in
form of dense, cold clouds and filaments.  The majority of the gas
lifted up by the fountains falls back on the disk remaining within a
radial distance $\Delta R=0.5$ kpc from the place where the fountain
originated. This localized circulation of disk gas does not
  influence the radial chemical gradients on large scale, as required
  by the chemical models of the Milky Way which reproduce the
  metallicity distribution without invoking large fluxes of metals.
Simulations of multiple fountains fuelled by Type II supernovae of
different OB associations will be presented in a companion paper.

%Our adaptive mesh scheme allow us to 
%follow in detail the limited volume where the supernovas explode and the extended 
%area where the ejected gas spreads.

\end{abstract}

\begin{keywords}

\end{keywords}

\section{Introduction} 
 \label{sec:introduction}
Deep H {\sevensize I} observations and a detailed modelling of the H
{\sevensize I} in edge-on spiral galaxies as NGC 891
\citep{swsava97,oofrsa07}, UGC 7321 \citep{mawo03} or partially
inclined spiral galaxies as NGC 2403 \citep{scsasw00,frat02}, NGC 4559
\citep{bar05} have shown the presence of thick gaseous layers
extending to several kiloparsec from the plane of the disk, with a
mass of 1/10 of the total H {\sevensize I}. Optical observations have
also revealed extraplanar emission from diffuse ionised gas (DIG)
\citep{fewyga96,howara99} which, as the H {\sevensize I}, extends to a
large extent above the plane.
The extraplanar gas (EPG) has peculiar kinematics that has been
studied both for the neutral and the ionised component. The modelling
of the H {\sevensize I} layer has shown that the EPG rotates in the
same direction as the disk, and has a mean rotation velocity about
20-50 km s$^{-1}$ lower than that of the gas on the plane \citep
[e.g.][]{swsava97,froo04}. Similar results hold for the ionised
component \citep{fra04}.

In general, the extraplanar gas appears to be regularly distributed
over the whole galaxy, and its amount is related to the rate of star
formation \citep [e.g.]{scsasw00,det05}. This suggests that a
significant fraction of the EPG has an internal origin, likely being
disk ISM ejected in the halo by (clustered) supernova explosions. 

The presence of giant holes in the H {\sevensize I} disk of many
galaxies as NGC 6946 \citep{boom05}, NGC 2403 \citep{frat02}, M33
\citep{deuden90}, M 101 \citep{kasava91}, M31 \citep{bribaj86}, and in
the Milky Way by \citep{hei84} and \citep{kohere92} also point to an
internal origin of the EPG. These holes can be hardly created by the
competing scenario of infalling clouds \citep{com04}. Instead, they
can easily accounted for by associations of O and B stars exploding in
a correlated fashion and giving rise to superbubbles. In some cases
vertical motions are associated with these holes, as expected once the
superbubble breaks out of the disk \citep{mamcno89,kmck92},
reinforcing the hypothesis that they are linked to galactic fountains.

The above arguments point at a substantial gas circulation between
disk and halo. The ejection of the gas out of the disk is thought to
be related to star formation in the disk and to be due mainly to Type
II supernovae (SNe II). This process is usually described as {\it
  galactic fountain} \citep{shafi76, breg80,kahn81}.  In the simple
version, the scenario assumes that gas heated by SNe explosions rises
above the plane of the disk but remains bound to the galaxy. Then, the
material will recirculate, falling backward to the galactic plane as
cold gas, presumably concentrated in clouds formed through thermal
instabilities.  According to this model, the galactic fountain gas
moves to larger galactocentric distances as it rises above the
galactic plane, to conserve its angular momentum.  This effect would
account for the observed lower rotation velocity of the extraplanar
gas.  When the clouds form via thermal instability, they fall back
toward the galactic plane and their radial distance from the centre
decreases. From a simple analysis it can be shown that after about $10^8$ yr 
of its formation a cloud returns to the Galactic plane at a Galactocentric 
radius not very different from its initial value (Spitzer 1990).
In the framework of a one-dimensional model \citet{kah93} showed analytically 
that, at the Sun distance from the Galactic centre, a cloud collide with the 
Galactic midplane at a distance of 850 pc from its initial position.
In these admittedly idealized scenarios, one expects the early outflow phase 
to be mostly traced by ionised gas and the late inflow phase by neutral gas.

Despite the arguments reckoned above for an internal origin of most
extraplanar gas, a scenario in which the halo cold and warm gas is
accreted from the intergalactic medium is not ruled out
\citep[e.g.][]{bli99}.  In fact, the above analysis could be reversed
to conclude that a higher star formation rate might be the final
result of accretion rather than the cause at the origin of the EPG.
Indeed, extended infall of external gas on galactic disks is needed in
current galactic evolution models as a way to maintain star formation
rates similar to the observed present ones in normal spirals \citep
[e.g.][]{takhir00,semcom02}, as well as to explain the observed
patterns in the chemical evolution of the Milky Way
\citep{chi02,gei02}.  The observational counterpart of this gas
accretion might be the high velocity clouds (HVCs) observed in our and
other galaxies \citep{wak01,wak04}. However, the ability of these
clouds to provide the needed amount of gas at low metallicity, as well
as their capability to trigger star formation is still unclear.
Evidence for star formation induced by HVC impacts on galactic disks
is unclear, while numerical simulations seem to indicate that
infall-induced star formation is substantially prevented
\citep[][and references therein]{com04}.

The galactic fountain scenario can also explain the diffuse soft
($\la$ 1 keV) X-ray emission that has been detected in some spiral galaxies
like M101 \citep{snpi95}, M51 \citep{ehpibe95}, M33 \citep{lochbl96},
NGC 891 \citep{brho97}, NGC 4631 \citep{waetal01}, NGC 2403
\citep{fretal02}, NGC 3556 \citep{wachir03}, and NGC 2613
\citep{lietal06} (for a short review see \citet{wa07}). This hot gas
does not extend significantly more than 10 kpc away from the galactic
disks, and appears to have substantial substructures which are best
appreciated in disk galaxies that are moderately inclined. In NGC 2841
the diffuse X-ray emission arises primarily from the inner galactic
disk, and its morphology is the result of various hot plumes sticking
out of the disk and into the halo of the galaxy \citep{wa07}.  These
X-ray emitting plumes extend vertically up to a few kiloparsecs and
are most likely blown out hot gas heated in massive star forming
regions in the disk. This conclusion is reinforced by the observation
that the X-ray luminosity correlates with the star formation rate and
that the hot gas seems oxygen rich, a signature of SNII enrichment.

From all the above, it turns out the fountains may be crucial in
characterizing X-ray, as well as H {\sevensize I} and H$\alpha$
emissions in normal spiral galaxies. Moreover galactic fountains could
also play a role in the chemical evolution of spiral galaxies,
especially in shaping the abundance gradients in the disk.  In
principle the fountains might be able to move SN II ejecta
relatively far from the place where they have been produced, affecting
the gradients of $\alpha$-elements such as Mg and O.

The richness of phenomena involved in the galactic fountains can be
hardly taken into account by the simple scenario outlined above
\citep{shafi76,breg80,kahn81}.  
Nearly twenty years after these three seminal papers, however, 
the advent of powerful computers made
possible to simulate the fountains rather accurately.

In fact, extensive work on superbubble/chimneys and fountains
formation and on their effects upon the interstellar medium in the
disk and the halo has been performed over the last decades \citep [e.g.]
[]{tomi86,norike89,maclow88,mamcno89,houck90,kor99} 
most of which has shown 2D and 
3D simulations considering the evolution of bubbles and superbubbles 
both in unmagnetized and magnetized media. 
In particular, \citet{teno88} and \citet{norike89} 
\citep [see also][] {tomi86,mamcno89}, 
were the first to explore the role that superbubles
and chimneys physically connected to underlying OB associations in
the disk play on the fountains evolution in galaxies. \citet{houck90} 
borrowig concepts from earlier works, investigated
the physical conditions for the formation of subsonic and transonic
fountains as a function of the radiative cooling time.

More recently, \citet{deav00} and \citet{deav01} performed a number of 3D
hydrodynamical simulations of the gas in the Milky Way in order to
account for the collective effects of supernovae on the structure of
the interstellar medium (ISM). In order to obtain a high spatial
resolution, only a small region of the Galaxy has been considered by the
authors; the simulated volume has an area of 1 kpc$^2$ and a vertical
extension varying between 4 and 10 kpc. The gas in this volume is
initially assumed to be plane parallel stratified, and any dependence
of the variables on the Galactocentric distance is neglected. The
general structure of the ISM is reproduced, with a thin H {\sevensize
I} disk overlaid by thick H {\sevensize I} and H {\sevensize II} gas
disk. This structure is essentially given by SNe II isolated or
assembled in small clusters.  The gas powered by SNe II associated in
larger clusters reaches larger heights, where it cools forming clouds
which eventually descend. Intermediate velocity clouds (IVCs) form
preferentially in the range $0.8<z<2$ kpc whereas HVCs (10\% of total
number of clouds) form at larger heights.

Three-dimensional simulations have been reported also by \citet {kor99}
who include the galactic magnetic field in the disk and the Galactic
differential rotation, but the vertical extension of the simulated
volume was $\pm$ 1 kpc, covering an area of 500 by 500 pc$^2$ with a
resolution of 8 pc. The lack of an extended grid made the development
of a disk-halo-disk cycle impossible, because part of the mass is
lost through the grid boundaries. MHD simulations were carried out also 
by \citet{deav05}, without including differential rotation, where 
again in order to reach very high resolution (of 0.6 pc) they have 
considered only a small volume of the Galaxy. 
These MHD studies have revealed that the gas
transport into the halo is not prevented by the parallel magnetic
field of the Galaxy, but only delayed by few tens of Myr when compared to 
pure HD simulations. 

\citet{frbi06} followed a different strategy. They considered the
whole Galaxy, and calculated orbits of gas clouds depicted as bullets
that are ejected (by SNe II) from the disk and then move ballistically
up into the halo and back down to the disk. Hydrodynamical effects are
essentially neglected in these models.

In this paper we would like to study the motion of the gas on large
scales in order to understand the observed kinematic of the EPG and
the possible influence in shaping the metallicity gradient in the
disk.  To this aim, we need different kind of simulations with respect
to the aforementioned papers \citep{kor99,deav00,deav01}.  In fact,
although \citet{kor99} take into account the Galactic rotation, the
simulated volume is to small to follow the gas circulation over large
distances.  The volumes considered by \citet{deav00} and
\citet{deav01} are larger, but in this case the rotation is neglected
and any radial dependence is absent.  On the other hand,
\citet{frbi06} take into account the whole Galaxy, but their approach
is purely ballistic.  Our approach is to consider the whole Galaxy and
run 3D hydrodynamical simulations of the fountains. While the
numerical grid used here includes the whole rotating Galaxy, our
adaptive mesh scheme allow us to follow in detail only a limited
volume of it, the remaining galactic volume being mapped at a lower
resolution.  Although a detailed description of the different phases
of the ISM is hampered by this limited resolution, the large scale
dynamic and thermal evolution of the fountains can be satisfactorily
followed. Here we present simulations of single fountains, i.e.
generated by a single OB association. We shall describe simulations of
multiple fountains in a companion paper.

\section{the model} 
 \label{sec:models}
\subsection{Galaxy model}
 \label{subsec:galmod}
The ISM in our model is made up of three components, namely molecular
(H$_2$), neutral (H {\sevensize I}) and ionized (H {\sevensize II})
hydrogen. Following \citet{wolf03}, each density component in
the disk is assumed to be of the form:
\begin{equation}
\label{eq:ism}
\rho = {\Sigma_{\rm d} \over 2z_{\rm d}}\exp\left(-{R_{\rm m} \over R}-
{R\over R_{\rm d}}- {|z| \over z_{\rm d}}\right ),
\end{equation}
\noindent
where $R$ is the cylindrical radius, $z_{\rm d}$ is the vertical scale
height and $R_{\rm d}$ is the radial scale length of the disk. The
parameter $R_{\rm m}$ allows for the depression in the gas density
observed in the inner several kiloparsecs of the Galaxy.  $\Sigma_{\rm
d}$ represents the superficial density of the different components.
The values of the parameters are summarized in Table \ref{tab:ism}.

The ISM is initially set in rotational equilibrium in the Galactic
gravitational potential given by the summation of dark matter halo, bulge and
disk contributions.

The dark matter halo gravitational potential is assumed to follow the
Navarro, Frenk and White profile \citep{nfw96}
\begin{equation}
\label{eq:phiblg}
\Phi_{\rm dm} (r)=-4\pi G r^2_{\rm dm,0}\rho_{\rm dm,0}{\ln(1+x) \over x},
\end{equation}
\noindent
where $\rho_{\rm dm,0}$ is a reference density, $r_{\rm dm,0}$ is a
scale radius, $x=r/r_{\rm dm,0}$ and $r$ is the spherical radius. The
halo is truncated at a radius $r_{\rm dm,t}$ beyond which its
potential follows the $1/r$ profile. See Table 2 for the numerical values
of these parameters.

The bulge gravitational potential is given by \citep{her90}:
\begin{equation}
\label{eq:phiblg}
\Phi_{\rm b} (r)=-{GM_{\rm b} \over {r_{\rm b,0}+r}},
\end{equation}
\noindent
where $r_{\rm b,0}$ is a scale radius and $M_{\rm b}$ is the bulge mass.  

Finally, the gravitational potential of the disk is assumed to be generated by a
stellar distribution following a flattened King profile
\begin{equation}
\label{eq:rhodisk}
\rho_{\star}(r)={\rho_{\star,0} \over \left [1+\left (R/ 
R_{\star,{\rm c}}\right)^2 +\left (z/z_{\star,{\rm c}}\right)^2\right]^{3/2}}.
\end{equation}
\noindent
Here $\rho_{\star,0}$ is the central density of the stars, and
$R_{\star,{\rm c}}$ and $z_{\star,{\rm c}}$ are the core radii, whose
ratio is $\delta=z_{\star,{\rm c}}/R_{\star,{\rm c}}=0.03$ In order to
avoid an unbounded growth of the stellar mass with radius the stellar
profile is truncated wherever $\sqrt{(R/R_{\star,{\rm
c}})^2+(z/z_{\star,{\rm c}})^2}> R_{\star,{\rm t}}/ R_{\star,{\rm
c}}\equiv z_{\star,{\rm t}}/z_{\star,{\rm c}}$, with $R_{\star,{\rm
t}}$ and $z_{\star,{\rm t}}$ tidal lengths whose ratio is
$z_{\star,{\rm t}}/R_{\star,{\rm t}}=\delta$. The stellar potential
$\Phi _{\star}$ is computed numerically following the method described
in \citet{brma96}.

Table \ref{tab:gal} gives the values of all the parameters concerning
the Galaxy model. With such values we obtain $M_{\rm dm}=1.3\times
10^{12}$ M$_{\odot}$, $M_{\rm b}=3.5\times 10^{10}$ M$_{\odot}$ and
$M_{\star}=9.6\times 10^{10}$ M$_{\odot}$ for the dark matter halo, the bulge
and the stellar disk, respectively.

In order to set the ISM given by Eq. \ref{eq:ism} in a rotating
configuration in equilibrium in the general potential well of the
Galaxy, we proceed as follows. The pressure at any point is found
integrating the $z$-component of the hydrostatic equilibrium equation
for any value of the disk radius $R$.  The integration starts at the
outermost values of $z$ (where we can assume the pressure $P=0$) and
proceeds inward to reach the galactic plane $z=0$.

\begin{figure}    
\begin{center}    
\psfig{figure=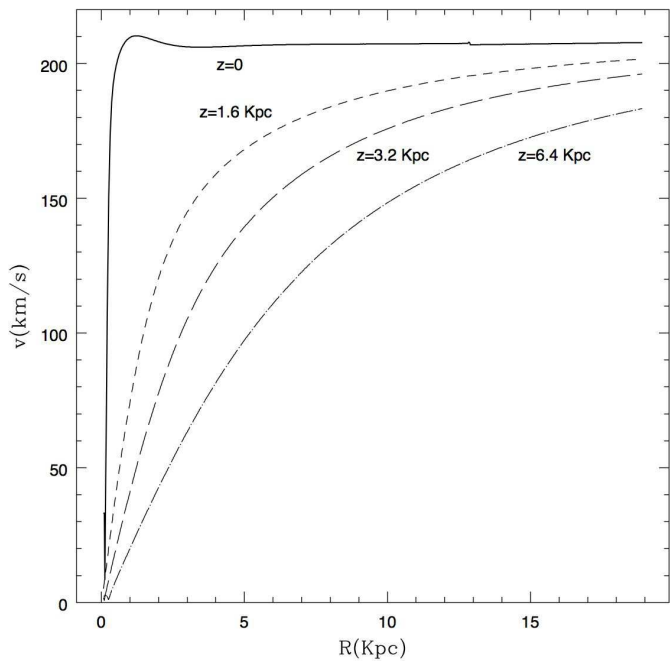,width=0.5\textwidth}    
\end{center}   
\caption{
Galactic rotation curves for different values of $z$.
}
\label{fig:rotcu} 
\end{figure}

\begin{figure}    
\begin{center}    
\psfig{figure=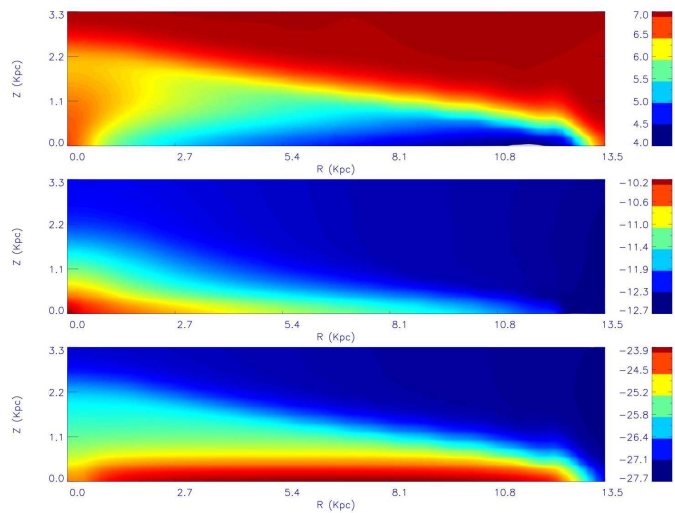,width=0.5\textwidth}    
\end{center}   
\caption{
Edge-on logarithmic distribution of temperature (top panel), pressure (middle
panel) and density (bottom panel) of the unperturbed ISM in our
Galaxy model. Distances are given in kpc, temperatures in K, pressure and density
in c.g.s.
}
\label{fig:gal} 
\end{figure}

We then obtain the rotation velocity from
\begin{equation}
v^2_{\phi}(R,z)=v^2_{\rm c}(R,z)-{R \over \rho}\left |{dP(R,z) 
\over dR }\right|_{z={\rm const.}},
\label{eq:vrot}
\end{equation}
\noindent
where $v_{\rm c}(R,z)=\sqrt {R(d\Phi(R,z) /dR)_{z={\rm cost}}}$ is the
circular velocity, and $\Phi$ is the total potential of the Galaxy:
$\Phi=\Phi_{\rm dm}+\Phi_{\rm b}+\Phi_{\star}$.  The values $v_{\phi}(R,0)$
obtained in this way mimic quite well the rotation curve of the Galactic
disk (c.f. Fig. \ref{fig:rotcu}).

\begin{table}
 \centering
 \begin{minipage}{140mm}
  \caption{ISM parameters.}
  \label{tab:ism}

  \begin{tabular}{@{}lcccc@{}}
%  \begin{tabular}{@{}llrrrrlrlr@{}}
  \hline
      & $\Sigma_{\rm d} $ & $R_{\rm m}$ & $R_{\rm d}$ & $z_{\rm d}$ \\
Components & $M_{\odot}$ pc$^{-2}$ & kpc & kpc  & pc \\
 \hline
 H {\sevensize I} ($R\leq 13$ kpc)   & 7.94 & 1.0 & 1000 & 178 \\
 H {\sevensize I} ($R\geq 13$ kpc)   & 571  & 10  & 4.0 & 324 \\
 H$_2$   & 57.5 & 3.3 & 2.89 & 63.4 \\
 H {\sevensize II}  & 1.39 & 0   & 30.0 & 880 \\
\hline
\end{tabular}
\end{minipage}
\end{table}

As shown in Fig. \ref{fig:gal}, that contains the temperature,
pressure and density distributions of the Galaxy, the gas has a
temperature always greater than $10^4$ K.  This comes from two
conditions. On one hand, from the necessity to hold in equilibrium the
disk gas distribution against gravity, as given by \citet{wolf03}.  On
the other hand, and more important, it is due to the assumption that H
{\sevensize I} has a smooth (rather than a clumpy) distribution, so
that a thermal, rather than a turbulent, pressure is required to
balance gravity.

In most models a hot ($T_{\rm h}=7\times 10^6$ K) isothermal gas halo
is added which is in equilibrium with the Galactic potential well:
\begin{equation}
\rho_{\rm h}=\rho_{0,{\rm h}}\exp[-(\Phi(R,z)-e^2\Phi(R,0)-(1-e^2)\Phi(0,0))/c^2].
\end{equation}
\noindent
Here $\rho_0=2.17\times 10^{-27}$ g cm$^{-3}$ is the halo central
density, and $c$ is its isothermal sound speed. In some models the hot
halo is allowed to rotate with velocity $v_{\phi,{\rm h}}=ev_{\rm
  c}(R,0)$, where $e$ is an arbitrary parameter with values in the
range $0\leq e \leq 1$.  In the models where the halo is present, the
disk density is replaced by the hot halo density wherever the halo
pressure is larger than the pressure of the disk. When this halo is
included, the initial pressure values which are needed to integrate the
$z$-component of the hydrostatic equilibrium equation (see above) are
given by the halo pressure at the $z$-boundary.

\subsection{Supernovae explosions}
 \label{subsec:snexp}
 It is known that the great majority of the massive stars form in
 clusters \citep[e.g.][and references therein]{watho02}. In the
 following, we define ``single fountain'' as a fountain powered by the
 SNe II present in a localized single cluster (for the sake of simplicity, with
 SNe II we mean here all the explosions due to core collapsed stars,
 i.e.  genuine SNe II, SNe Ib and SNe Ic). At the observed rate of 
 $1.2 \times 10^{-2}$ yr \citep{cap97}, several single fountains may
 occur sufficiently close to each other in time and space to interact
 mutually. This interaction obviously modifies the general dynamics of
 the SN II ejecta of a single fountain. In order to understand the
 effect of different factors (such as the Galactocentric distance, the
 presence of a gaseous halo, the rotation velocity of this halo), and
 how these effects are influenced by the interaction among several
 fountains, we have computed two classes of models differing only in
 the assumed number of stellar associations, i.e., single fountain
 models (SFMs) and multiple fountains models (MFMs). Here we focus on
 the first class of models, while the latter will be discussed in a
 forthcoming paper.

\subsubsection{Single association}
 \label{subsubsec:singlesnexp}

 In the case of a SFM, we assume that the OB association is
   located on the Galactic plane ($z=0$), and is totally housed
 inside a single mesh of the highest resolution level of our adaptive
 grid (cf. sect. \ref{sec:nummet}). The SNe II do not explode all at
 the same time because of the different masses of their
 progenitors. Thus the energy and the mass delivered by all the
 explosions are released over a time interval $\Delta t$=30 Myr (the
 lifetime of an 8 $M_{\odot}$ star, which is the least massive SN II
 ancestor).  We inject the explosion energy in the form of thermal
 energy at a rate $\dot E={\cal R}E_0$, where $E_0=10^{51}$ erg is the
 energy of a single explosion, ${\cal R}=N_{\rm SN}/\Delta t$ is the
 SN II rate, and $N_{\rm SN}$ is the total number of SN II explosions
 occurring in the association. The mass of the SN II ejecta is
 injected at a rate $\dot M ={\cal R}M_{\rm ej}$, where $M_{\rm
   ej}=16$ $M_{\odot}$ is the mean mass expelled by a single SN II
 explosion.  As a result, we do not consider a sequel of instantaneous
 explosions but, instead, a wind inflating a bubble with a mechanical
 luminosity $L_{\rm w}=\dot E$. This treatment is supported by
   the findings of \citet{maclow88} who showed that discrete
   explosions can be approximated as continuous events when more than
   $\sim 7$ SNe come one after the other.  About the temporal wind
   evolution, \citet{leith99} showed that SNe belonging to an
   istantaneous burst release mass and energy at rates oscillating
   around a mean value; the general properties of the fountain we are
   looking for depend on such mean values rather than on the
   oscillations. We thus assumed constant values of  $\dot M$ and $L_{\rm w}$,
as described above.

As the bubble expands, the swept up
 ISM is compressed in a dense radiative shell edging the bubble
 itself. The interior of the bubble, instead, is filled by the hot,
 low density ejecta \citep[c.f.][]{weav77}.
The bubble expands faster in the direction perpendicular to the plane,
because the density decreases more steeply in that direction. If the
density gradient is strong enough, the shell can accelerate. The
acceleration of the shell causes it to be disrupted by the
Rayleigh-Tailor instability, and the hot gas inside the bubble
subsequently leaks out and drives a new shock into the ambient medium,
giving rise to the galactic fountain.  The break out is expected to occur when
$L_{\rm w}>2.5L_{\rm b}$, where the critical luminosity is given by
\citep[c.f.][]{kmck92}.
\begin{equation}
\label{eq:lcr}
L_{\rm b}=2.4\times 10^{36}P_{0,4}H_{\rm eff,2}c_{0,6}\;\;{\rm erg\;s^{-1}.}
\end{equation}
\noindent
Here $P_{0,4}=P_0/(10^4\kappa$ cm$^{-3}$ K), where $P_0$ is the
midplane pressure, and $\kappa$ is the Boltzman's constant.
$c_{0,6}=c_0/(10^6$ cm s$^{-1}$), where $c_0$ represents the sound
speed in the midplane; actually, as the temperature varies somewhat
within the range $0<z<H_{\rm eff}$, we adopt the mean value of $c$
in Eq. \ref{eq:lcr}. Finally, the effective scale heigh of the
ISM at the Galactocentric distance $R$ is defined as
\begin{equation}
\label{eq:heff}
H_{\rm eff}={1 \over \rho(R,0)}\int_0^{\infty}\rho(R,z)dz
\end{equation}
\noindent
All the SFMs we run have the same number $N_{\rm SN}=100$ supernova
explosions. This number has been chosen because typical for Galactic
OB associations \citep[][see also Paper II]{hili05}. Furthermore, such
a number assures the realization of the break out at any location of
the fountain on the disk (cf. sect. \ref{subsec:galdis}).

It is worthwhile to note that our numerical grid is at rest in space,
while the Galaxy rotates. Thus, the OB association moves around
describing a circle on the $z=0$ plane.  In order to properly locate
the energy and mass source in the grid, we compute separately the
trajectory of the association, and at each time we put an appropriate
amount of thermal energy and mass in the smallest cell of our
  multilevel grid (see below) transited by the association at that
time.
 
\begin{table*}
 \centering
 \begin{minipage}{140mm}
  \caption{Galactic parameters.}
  \label{tab:gal}
  \begin{tabular}{@{}cccccccccc@{}}
%  \begin{tabular}{@{}llrrrrlrlr@{}}
  \hline
   $\rho_{\rm dm,0}$ & $\rho_{\star,0}$ & $M_{\rm b}$ & $r_{\rm dm,0}$ 
& $r_{\rm dm,t}$ & $r_{\rm b,0}$ & $R_{\star,{\rm t}}$ & $R_{\star,{\rm c}}$
& $z_{\star,{\rm t}}$ & $z_{\star,{\rm c}}$ \\
$10^{-24}$ g cm$^{-3}$ &  $10^{-24}$ g cm$^{-3}$ & $10^{10}$ $M_{\odot}$ &
kpc  &kpc  &kpc  & kpc & kpc & kpc & kpc \\
 \hline
 0.29 & 120 & 3.5 & 30.8 & 347.7 & 0.8 & 20 & 1.2 & 0.6 & 0.036 \\
\hline
\end{tabular}
\end{minipage}
\end{table*}

\section{The numerical method} 
\label{sec:nummet}

In our simulations we have used a modified version of the adaptive
mesh refinement YGUAZU code. This code integrates the 3D inviscid
gasdynamic equations (expressed in Cartesian coordinates) with the
flux vector splitting algorithm of \citet{leer82}. Radiative losses
are computed through a non-equilibrium ionization (NEI) calculation
including a set of continuity equations for atomic/ionic or chemical
species; in the present simulations solar abundances \citep{Aspl} are
assumed.  The details of the gasdynamic and the adaptive grid
algorithms have been presented by \citet{raga00} and \citet{Raga02},
and tests of the code, as well as, several physical applications can
be found in \citet{Raga02}, \citet{masc02}, \citet{gonz04} and
\citet{mel05}. When compared with other methods, this particular
scheme has a reasonable performance (as far as diffusion is concerned)
when one considers, e.g., 1D test problems (see, e.g.,
\citet{vanal82}; \citet{wood04}). Artificial viscosity is set to zero
in all our simulations.  As discussed in the previous Section, at any
time the mass and energy source terms due to the SNe are absent
everywhere but in the cell hosting the OB association.

The 3D binary, hierarchical computational grid is structured with a
base grid, and with a number of nested grids whose resolution doubles
going from one level to the next one.  In our most refined
simulations, we have adopted six grid levels (with the linear size of
the meshes of the most refined grid being 13 pc) covering a box with
physical dimensions of $26.4\times 26.4\times 6.6$ kpc.  The grid
being adaptive, in principle the maximum resolution should be set at
the boundary between the hot halo and the disk, where the density
gradient (along the $z$ direction) is very steep. This, however, would
represent an useless waste of cpu time because we are interested in
describing in particular the fountain which evolves in the selected
area of the disk. We thus enforce the maximum grid resolution only in
this area and in the volume above it, even if the density gradients in
the halo are mild. In fact, we want to follow the circulation and the
thermal history (i.e. the degree of radiative cooling) of the metals
expelled by the SNe II, as carefully as possible. For this reason we
have added three different tracers passively advected by the code
describing the disk gas, the halo gas and the SN II ejecta. In this
way, we can distinguish the evolutive history of each gas component
and understand their degree of mixing, as the fountain evolves.

Given the symmetry of the problem, our computational volume
encompasses only the ``upper'' half of the Galaxy. We thus have
outflow boundary conditions everywhere but at ``the bottom'' (i.e. the
$z=0$ plane) where reflecting boundaries are enforced. 

\section{The reference model RM}
 \label{sec:refmod}

 In this section we discuss extensively the properties of the RM, which
 we take as a reference model. In this model
 the gaseous halo is present but does not rotate, and the OB
 association is located at the galactocentric distance $R=8.5$ kpc. At
 this radius the transition between the disk and the hot halo occurs
 at $z=800$ pc. The critical luminosity is $L_{\rm b}=1.5\times
 10^{37}$ erg s$^{-1}$, well below the mechanical luminosity $L_{\rm
   w}=10^{38}$ erg s$^{-1}$ provided by the 100 SNe II powering the
 fountain. This model has been run with a maximum spatial resolution
 of 13 pc.

\begin{figure}    
\begin{center}    
\psfig{figure=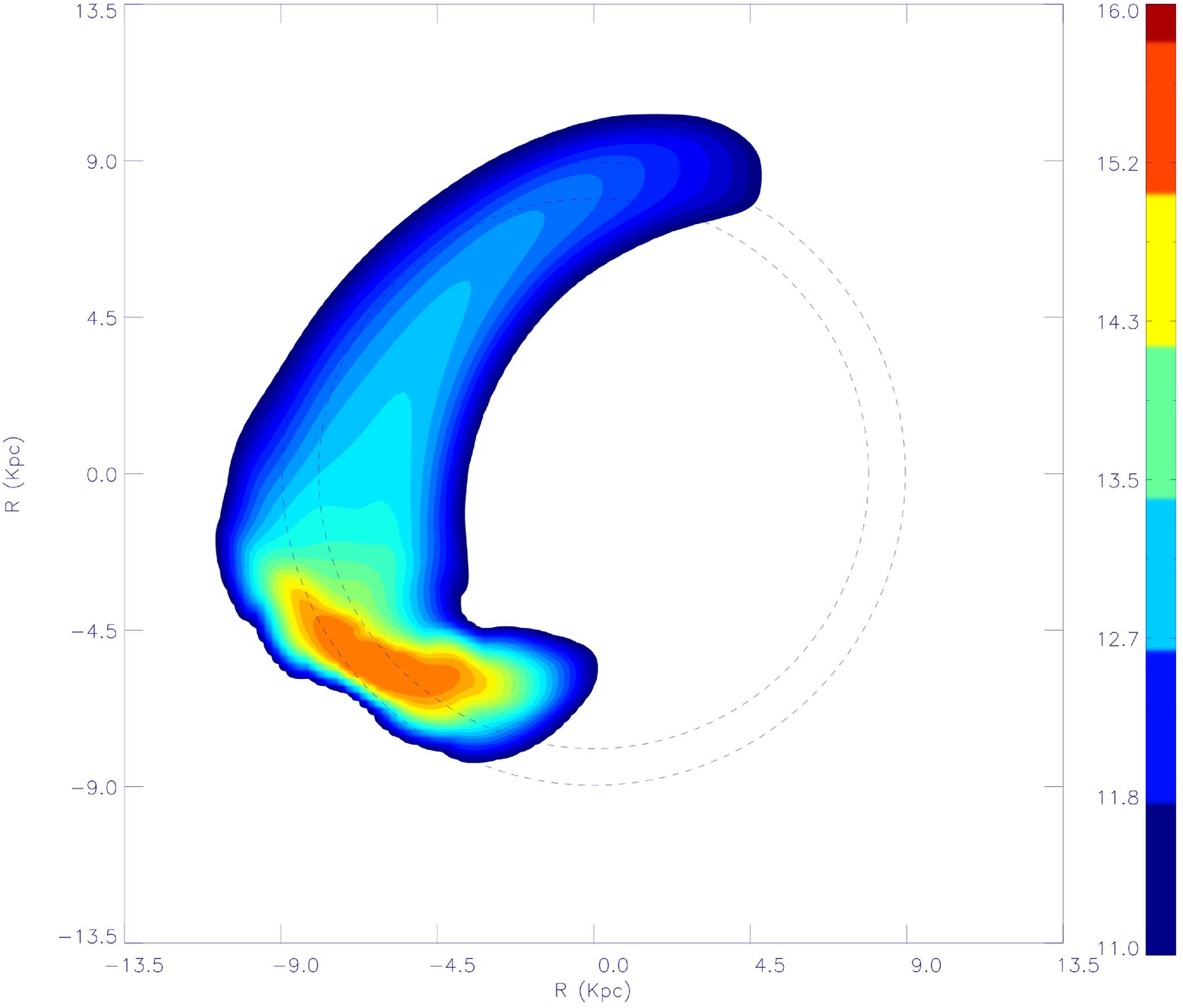,width=0.5\textwidth}    
\end{center}   
\caption{ Face-on view of the $z$ column density distribution of the
  SN II ejecta after $t=160$ Myr for RM. The SNe II explode at a radius
  $R=8.5$ kpc, halfway between the radii $R=8$ kpc and $R=9$ kpc of
  the two circles drawn in figure. These two circles are drawn to
  guide the eye, and their distance represents approximately the
  maximum extension of the hole in the ISM carved by the fountain on
  the disk during its activity. At the time to which the figure refers
  to the hole has collapsed and disappeared. The logarithmic column density 
  scale is given in cm$^{-2}$ and the $x-y$ scale is in kpc.  }
\label{fig:colden} 
\end{figure}
 
\begin{figure*}    
\begin{center}    
\psfig{figure=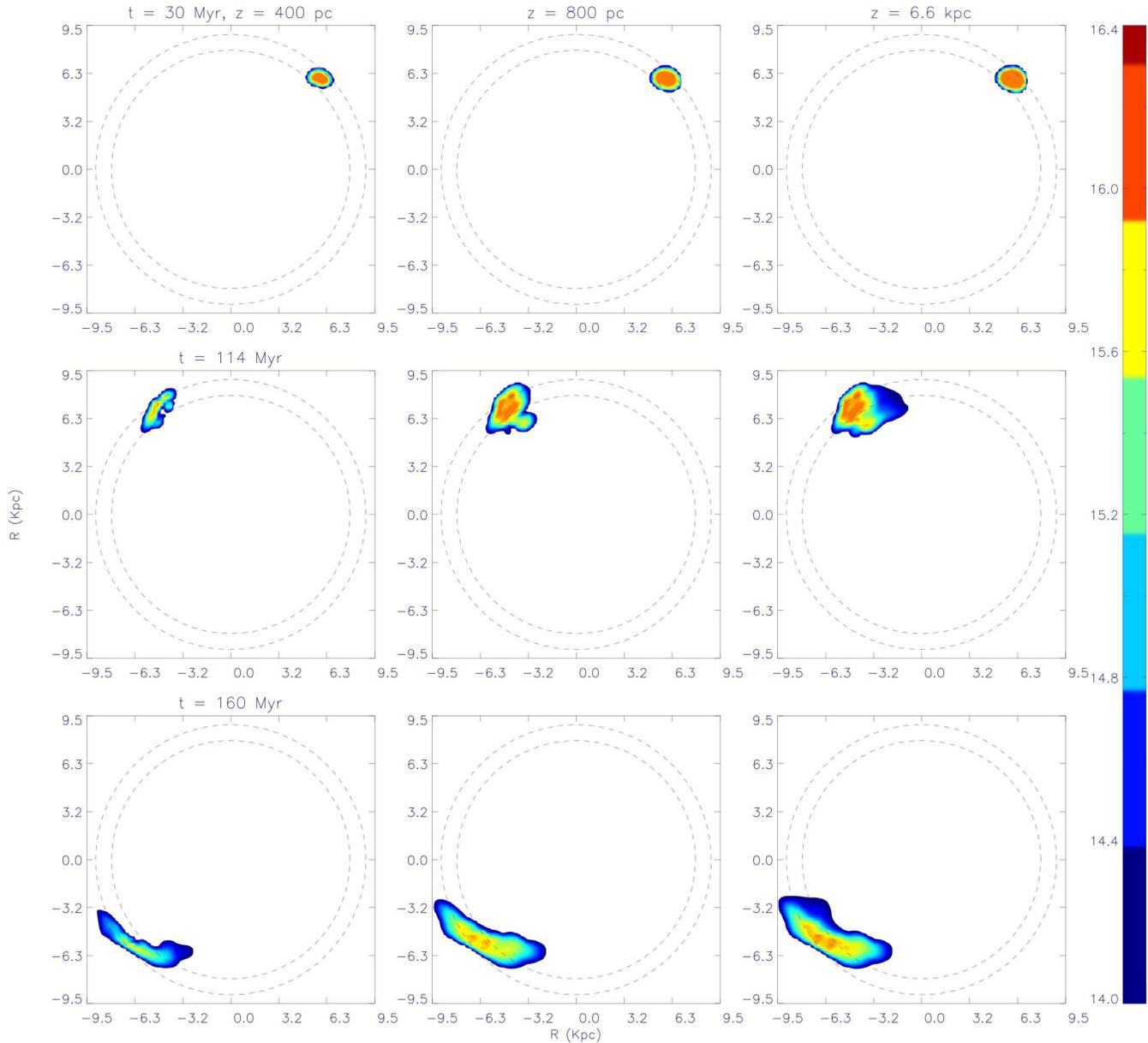,width=1.0\textwidth}    
\end{center}   
\caption{ Each row represents the face-on view distribution of the $z$
  column density of the SN II ejecta up to $z$=400 pc (left panel),
  $z$=800 pc (central panel), and $z=$6.6 kpc (right panel) for RM;
  the upper, middle and lower rows refer to $t=$30 Myr, $t$=114 Myr
  and $t$=160 Myr, respectively. The circles visible in the panels
  have the same meaning than in Fig.~\ref{fig:colden}. The logarithmic column
  density scale is given in cm$^{-2}$ and the $x-y$ scale is in kpc.
}
\label{fig:tomo} 
\end{figure*}

In Fig.~\ref{fig:colden}, we show the face-on column density of the SN
II ejecta after $t=160$ Myr. During its activity the fountain digs a
hole on the disk, and throws SN II ejecta and ISM vertically up to
$|z|=2$ kpc above the galactic plane. Once the stellar explosions
cease, the hole collapses in $\sim 2\times 10^7$ yr and the ejecta
trapped at its edges (nearly half of the total, see below) mixes with
the local ISM. Owing to the differential rotation of the Galactic
disk, the ejecta does not remain confined in one point, but is
stretched, giving rise to the bean-like structure seen in
Fig.~\ref{fig:colden}. The low density ``tail'' with a banana-shape is
given by the ejecta pushed at high altitude which then slowly falls
back lasting above the disk.

The two concentric circles in Fig.~\ref{fig:colden} with radii $R=8$ kpc
and $R=9$ kpc, have been drawn to guide the eye. The SNe II explode
between these two circles, and their distance corresponds to the
diameter of the hole produced by the fountain. Comparing the shape of
the tail traced by the ejecta with the circles, we note that the gas
of the fountain tends to move inward during its trajectory, rather
than outward, as one could expect, at least initially (cf. section
\ref{sec:introduction}).  Actually, as the gas moves upward and interacts
with the halo, it transfers to it part of its angular momentum; the
centrifugal force decreases in pace with the circular velocity and the
centripetal component of the gravity prevails, pushing the gas toward
the Galactic centre.

To better illustrate the fountain evolution, we show in
Fig.~\ref{fig:tomo} a sort of tomography of the fountain itself. Each
single row in this figure represents the column density of the ejecta
up to $z$=400 pc (left panel), $z$=800 pc (central panel), and $z>$6.6
kpc (right panel); the upper, middle and lower rows refer to $t=$30
Myr, $t$=114 Myr and $t$=160 Myr, respectively. From the first row, we
see that, initially the gas pushed by the fountain rises quite
vertically forming a column whose base has a size comparable to the
distance between the two circles. In the remaining panels the
progressive ``stretching'' of the ejecta is evident, as well as the
formation of the ``cometary'' tail at high values of $z$ and its
tendency to bend toward the Galactic centre.

\begin{figure}    
\begin{center}    
\psfig{figure=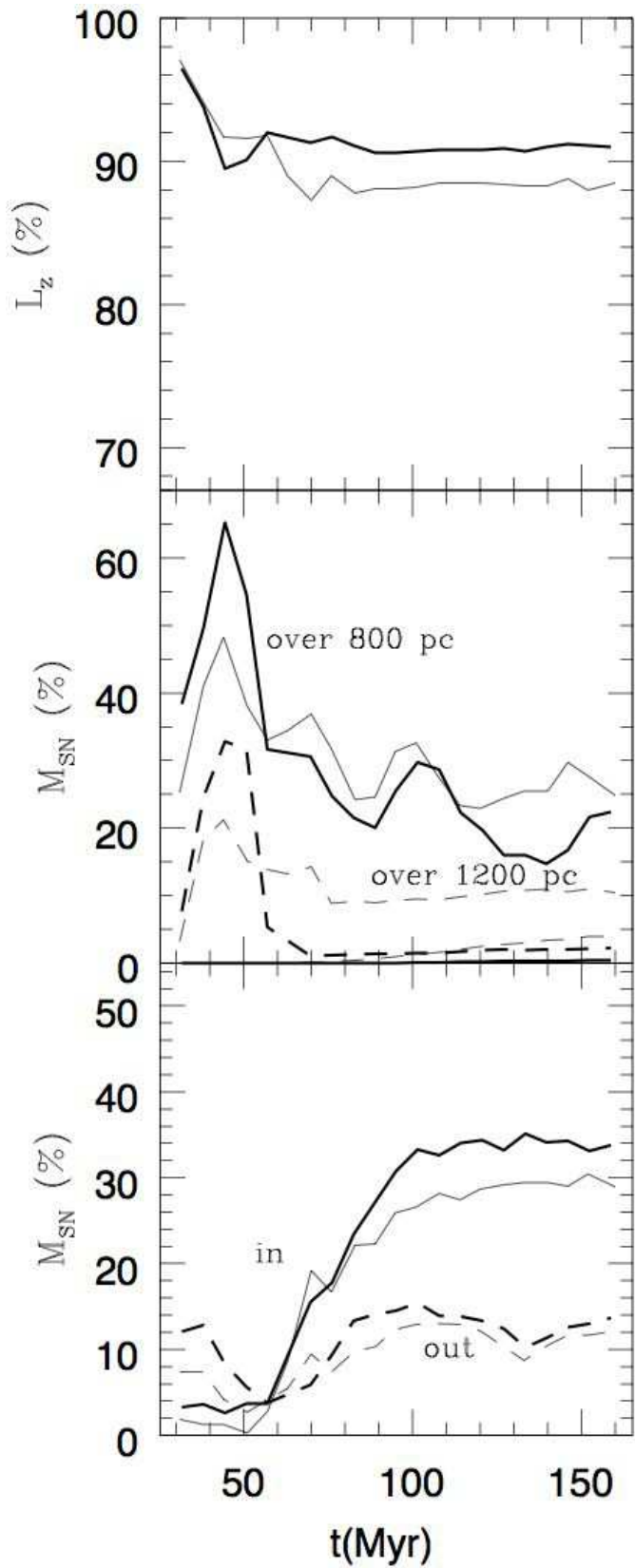,width=0.5\textwidth}    
\end{center}   
\caption{ The upper panel shows the evolution of angular momentum of
  the SN II ejecta for the RM. The evolution of the amount
  of the ejecta above different heighs is illustrated in the middle
  panel. Finally, the lower panel displays the temporal behaviour of
  the amount of the ejecta located at $R<8$ kpc (dashed line) and
  $R>$9 kpc (solid line); these two radii correspond to the circles
  visible in Figs~\ref{fig:colden} and ~\ref{fig:tomo}.  The thin lines
show the results obtained adopting a maximum resolution of 26 pc (see
text for the discussion}
\label{fig:angmo} 
\end{figure}

A further insight on the evolution of the galactic fountain is
obtained from Fig.~\ref{fig:angmo}. The three panels illustrate the
evolution of several quantities starting from $t$=30 Myr, the time at
which the SNe II stop to explode. The upper panel shows how quickly
the ejecta loses its angular momentum because of the interaction with
the gaseous halo. After 80 Myr nearly 10\% of the angular momentum of
the fountain has been transferred to the hot halo; later on, no
further transfer occurs because nearly 75\% fraction of the ejecta is
located below the disk-halo transition and rotates toghether with the
ISM \footnote{The transfer of angular momentum is evaluated comparing
  the angular momentum of the ejecta within the grid with the original
  angular momentum held by the ejecta at the moment of its injection
  within the grid.}.  This is illustrated by the middle panel of
Fig. \ref{fig:angmo} showing the amount of the ejecta mass fraction
located above different heights. It is interesting to note that, apart
for some fluctuations, for $t>$80 Myr, the long-term evolution of
these quantities becomes very slow. This is due to the fact that most
of the ejecta situated above the plane is rather diluted, and tends to
float together with the extra-planar ISM.  Finally, in the lower panel
of Fig.~\ref{fig:angmo} we quantify the tendency of the ejecta to move
radially by plotting the fraction of the total mass of ejecta located
at $R>$9 kpc and $R<$8 kpc (i.e. the radii of the two circles drawn in
Figs.~\ref{fig:colden} and ~\ref{fig:tomo}); the amount of mass
located within the region $8<R<9$ kpc is not taken into account. In
the beginning the ejecta starts to follow the expected tendency to
move outward but, as the loss of angular momentum proceeds, the
fraction of gas moving inward increases and after 60 Myr overrides
that directed outward. The amount of the outer mass remains always
very low and does not evolve very much; on the contrary, in the inner
region the mass of the ejecta keeps increasing during almost all the
period of the simulation.

In Fig. \ref{fig:angmo} are also shown results obtained with a
  coarser spatial resolution (of the finest grid) of 26 pc. Actually,
  in order to test the convergence of our simulations, we have run
  simulations with different higher refinement level resolution: 52,
  26 and 13 pc. The gross features of the hydrodynamical evolution of
  the fountain over the whole gas cycle do not change significantly
  and are captured even in the coarser case. We notice that in the
  coarser case there is a slightly larger deposition of material at
  higher latitudes (10\% larger) than in the higher resolution case
  due to smaller radiative cooling. As a consequence, the coarser case
  has also a slightly larger loss of angular momentum, however the
  overall behavior of the fountain evolution is found to be very
  similar in both cases and does not affect the main conclusions of
  the present work.

\begin{figure}    
\begin{center}    
\psfig{figure=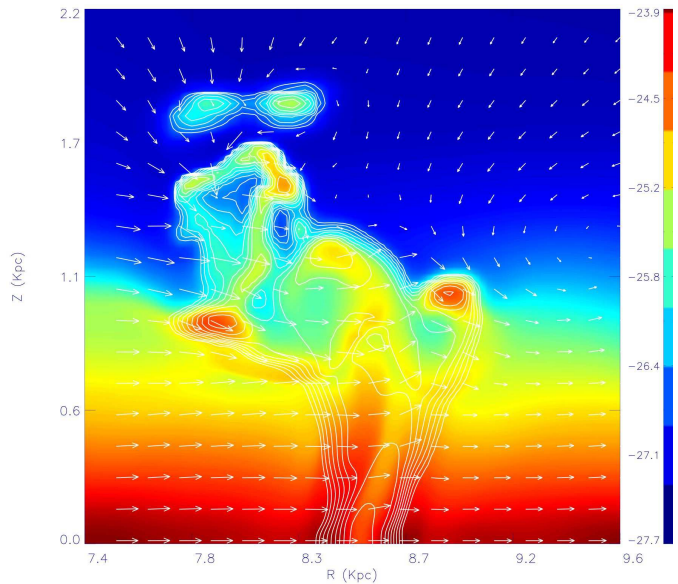,width=0.5\textwidth}
\end{center}   
\caption{Vertical logarithmic distribution of the gas at $t=50$
  Myr on a plane orthogonal to the Galactic plane and passing through
  the place where the SN II association is located. 
Logarithmic isodensity contours (g cm$^{-3}$) of the ejecta are 
overimposed, highlighting the fountain pattern and the cloud formation: 
the ten contours cover the range between -28.7 and -26.7, with
constant intervals of 0.2.
As the plane in figure does not encompass the Galactic
axis, the gas disk shows an apparent outward radial velocity given by the 
projection of its rotational velocity. 
This projection corresponds to 75 km s$^{-1}$}
\label{fig:cloud} 
\end{figure}

\begin{figure}    
\begin{center}    
\psfig{figure=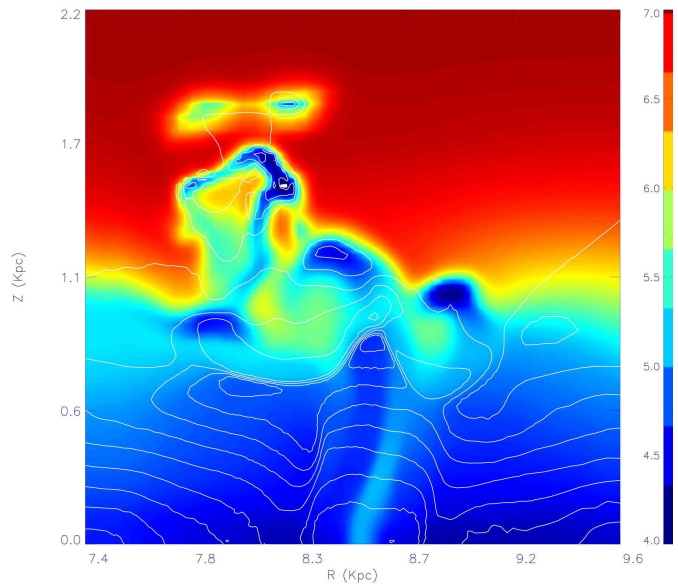,width=0.5\textwidth}    
\end{center}   
\caption{Vertical logarithmic distribution of the gas temperature at $t=50$
  Myr. Logarithmic isopressure contours (dyn cm$^{-2}$) of the pressure are 
  overimposed: the fifteen contours cover the range between -13. and -10.8, 
  with constant intervals of 0.146.}
\label{fig:cloud1} 
\end{figure}

The gas lifted up by the fountain has a mass $2.5\times 10^5$ $\msun$,
almost all (92\%) condensed into dense filaments cooled to $T=10^4$ K,
the minimum allowed temperature (see Fig. \ref{fig:cloud} and Fig.
\ref{fig:cloud1}).  The clouds form via thermal instabilities at
$z\sim 2$ kpc once the ascending gas has inverted its motion and moves
back toward the disk.  In fact, the clouds have all negative
$z$-velocities in the range 50-100 km s$^{-1}$. We point out that the
clouds in our simulations cannot form by the hot wind gas powered by
the SNe II. In fact, the sound crossing time of this gas is shorter
than the cooling time, and cold condensations are prevented
\citep{houck90}.  Instead, the clouds originate by the compression and
cooling of the disk gas lifted by the fountain.

%Along the first 30 Myr the disk gas sweept by the SNe shock 
%waves has a mass of $\sim$ $2 \times 10^6$ M$_{\odot}$, while the SNe 
%ejected gas has a mass of only $\sim$ 800 M$_{\odot}$. Due to this
The chemical composition of these clouds is practically
unaffected by the SN II ejecta. The fountain is powered by 100 SNe II,
half of them exploding in the half space mapped by the grid; as each
supernova delivers on average 3 $\msun$ of metals
\citep[cf.][]{mar07}, a total mass of 150 $\msun$ of heavy elements is
ejected by the fountain within the computational domain. At $t=50$ Myr,
when the lifted ISM reaches the maximum height,
50\% of these metals results trapped within the disk, while $\sim$
30\% remains diffuse floating over the disk as hot, diffuse gas
and only $\sim20$\% is
locked in the clouds. As a result, the metallicity increment
in the clouds due to the freshly delivered metals 
corresponds to $\sim 0.01$ in solar units, negligible compared to
the $\sim$ solar abundance of the ISM.
%%is $1.2\times
%%10^{-4}$, substantially negligible compared to the solar metallicity
%%of the ISM.
As we will show in a forthcoming paper, this result also
holds in models with multiple SN II associations.
In conclusion, almost all the gas lifted up by the fountain condense
into clouds without being chemically affected. 
After 150 Myr 45\% of the fresh metals stays ``on the disk'' 
(below $z=800$ pc) within a radial distance of
$\Delta R=0.5$ kpc from the OB association. A further fraction of 35\%
is found on the disk within the range 9.5$<R<$7 kpc. The remaining 20\% of
metals is still over the disk, half of it at $R>8.5$ kpc and half at
$R<8.5$ kpc.
(cf. Fig. \ref{fig:colden} and Fig. \ref{fig:tomo}).

\section{Changing the parameters}
 \label{sec:chparam}

 In this section, we illustrate the differences among the results
 obtained with our reference model and those taking
 different parameters, the aim being to understand the role played by these
 parameters in the evolution of a galactic fountain. In order
 to obtain a relatively large number of models, we have run them adopting a
 spatial resolution  for the finest grid (26 pc) which was worse than that
 adopted for the reference model (13 pc). This latter model was also
 run with this coarser grid in order to make homogeneous comparisons.

Although the results depend on the adopted spatial resolution, several
general characteristics of RM are not greatly affected by the minimum
mesh-size adopted, as shown in the top and bottom panels of
Fig. \ref{fig:angmo}. In the middle panel, instead, we note that the
peaks present at $t=40$ Myr are higher, while the successive
``plateau'' are lower, the difference being grater for the dashed
lines. The higher peaks are due to the smaller volume from which the
SN II energy is delivered: in the finer grid the same energy is in a
smaller volume and the resulting pressure pushing the gas is
larger. The lower plateau derives by the improved resolution with
which the clouds are followed by the finer grid: in this case, in
fact, clouds are smaller and denser, and can fall faster toward the
disk. These differences, however, will not influence the relative
comparison below among models.

\subsection{Varing the Galactocentric distance}
\label{subsec:galdis}

\begin{figure*}    

\begin{center}    
\psfig{figure=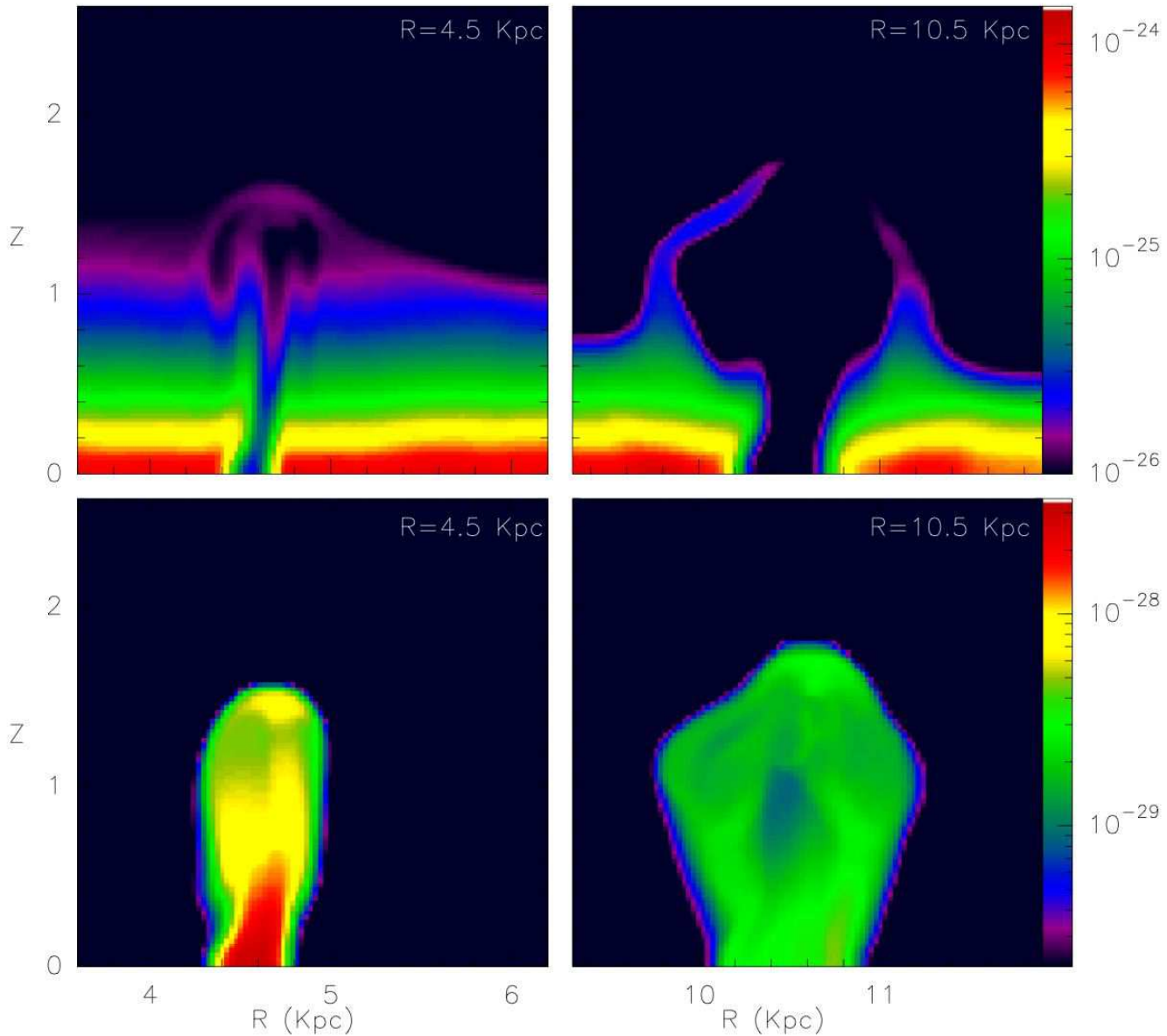,width=1.0\textwidth}    
\end{center}   
\caption{Edge-on view of two single galactic fountain models. The
  upper panels illustrate the cavities carved into the ISM by the
  fountains at $R=4.5$ kpc (left) and $R=10.5$ kpc (right) after $t=30$
  Myr. The lower panels show the ejecta distributions in the two
  cases. Distances are in kpc, and the c.g.s. density and ejecta scales are
 logarithmic.  }
\label{fig:r1r2} 
\end{figure*}

\begin{figure}    
\begin{center}    
\psfig{figure=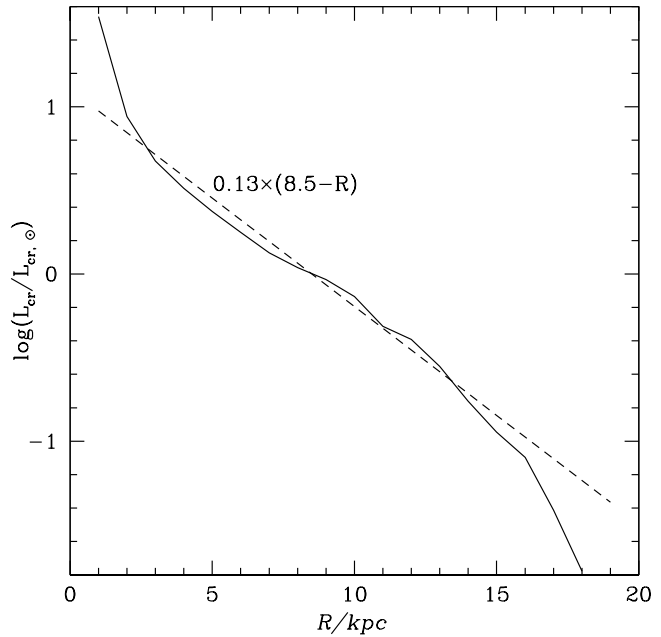,width=0.5\textwidth}    
\end{center}   
\caption{Variation of the critical luminosity with the Galactocentric
  distance, normalized to $L_{\rm cr,\odot}=1.5\times 10^{37}$ erg s$^{-1}$,
  the value of the critical luminosity at the solar position $R=8.4$
  kpc. The line represents an useful analytical approximation in the range
of interest.  }
\label{fig:lcr} 
\end{figure}

In this section, we explore the influence of the galactocentric
distance on the fountain dynamics. In particular, we run two models
similar to RM, but located at $R=4.5$ kpc (the inner fountain) and
$R=10.5$ kpc (the outer fountain). At these distances, the
corresponding critical luminosities
(cf. sect. \ref{subsubsec:singlesnexp}) are $L_{\rm b}=4\times
10^{37}$ erg s$^{-1}$ and $L_{\rm b}=10^{37}$ erg s$^{-1}$,
respectively (the general dependence of the critical luminosity on the
Galactocentric distance is given in Fig. \ref{fig:lcr}).

Figure \ref{fig:r1r2} illustrates the cavities carved into the ISM by
these two fountains (upper panels), as well as the ejecta distribution,
at $t=30$ Myr. Both ``chimneys'' show a tendency to bend toward
larger $R$ because of the Galactic rotation. While the outer fountain
experiences a well defined break out, the inner fountain barely crosses
the disk-halo transition because $H_{\rm eff}$ is larger at $R=4.5$ kpc.
The differences of the effective scale heights, together with the
differences of $L_{\rm b}$, are also responsible for the different
shapes of the two fountains. Following \citet{kmck92}, it is

\begin{equation}
\label{eq:rh}
{R_{\rm eq} \over H_{\rm eff}}=\left ({17.9 \over 2\pi}\right )^{1/2}
\left ({L_{\rm w} \over mL_{\rm b}}\right )^{1/2},
\end{equation}

\noindent
where $m$ is the Mach number of the wind relative to the ambient
medium, and $R_{\rm eq}$ is the radius of the reflected shock
(cf. \citet{weav77}); the contact discontinuity separating the shocked
wind from the ambient medium is at somewhat larger radius.
If $R_{\rm eq}/H_{\rm eff}$ is greater than unity, than it is possible
for the wind to escape freely along the vertical axis without being well
collimated. If instead $R_{\rm eq}\la H_{\rm eff}$ a collimated jet may form.
This is actually the case for the inner fountain where, for $m=10$,
$R_{\rm eq}/H_{\rm eff}=0.85$, while $R_{\rm eq}/H_{\rm eff}=1.8$ holds
for the outer fountain.

\begin{figure}    
\begin{center}    
\psfig{figure=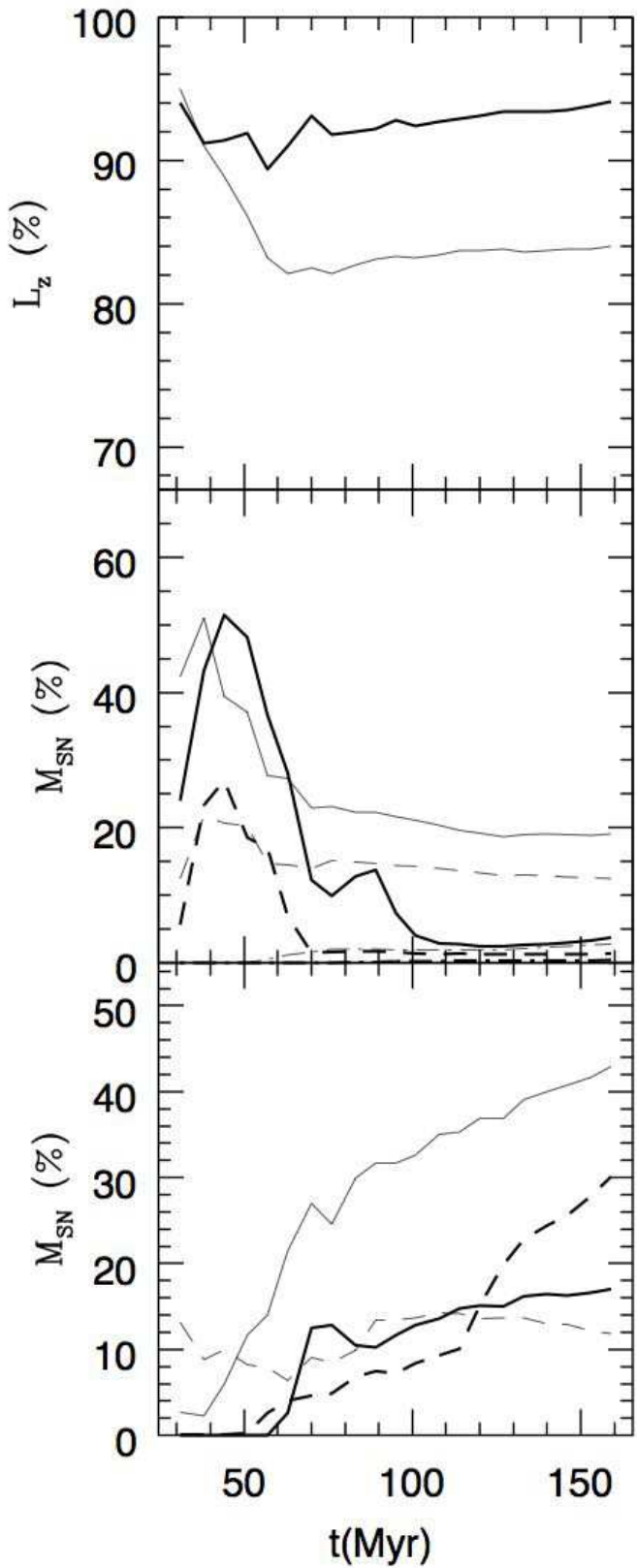,width=0.5\textwidth}    
\end{center}   
\caption{ Evolution of two models of fountain powered by the same
number of SNe II as in RM, but located at $R=4.5$ kpc (thick lines)
and $R=10.5$ kpc (thin lines) The meaning of the panels, as well as
that of the line types is the same as in Fig. \ref{fig:angmo} }
\label{fig:comrad} 
\end{figure}

More insight on the two fountains is given by Fig.~\ref{fig:comrad}.
The upper panel shows the evolution of the
angular momentum of the SN II ejecta of the two fountains. The loss of
angular momentum clearly increases with increasing values of $R$. To
understand this behaviour we must consider the different ambient medium in
which the fountains evolve and the different circulation of their
ejecta. This is summarized in the second panel of
Fig.~\ref{fig:comrad}. Up to 60 Myr all the fountains push the gas
essentially to the same heights (cf. also the middle panels of
Fig. \ref{fig:angmo}) despite the different values of $L_{\rm b}$.
The disk-halo transitions at $R=4.5$ kpc and $R=10.5$ kpc occur at
$z=1.35$ kpc and $z=0.45$ kpc, respectively. This means that most of
the ejecta of the inner fountain never reaches the hot halo, while the
opposite is true for the outer fountain (Fig. \ref{fig:r1r2}).

Once the expanding phase of the evolution ceases, the ejecta of the
inner fountain falls back almost entirely to the midplane under the action
of the surrounding pressure of the disk. In the case of the outer
fountain, instead, the fraction of the ejecta that reached the hot
halo falls back rather slowly, and ''fluctuates'' above the disk as in
RM.

As anticipated above, the different dynamical behaviour of the ejecta
influences its loss of angular momentum. In the case of the inner
fountain, the ejecta does not interact substantially with the hot halo
and suffers only a small reduction of its angular momentum; moreover,
as the ejecta is pushed back to the midplane, it is dragged by the
disk and its angular momentum increases.  Besides the fact that the
ejecta of the outer fountain emerges from the disk with a larger
angular momentum (per unit mass), it also loses a larger amount of it
because its interaction with the hot gaseous halo lasts longer and
involves a larger fraction of its mass.

Finally, the bottom panel of Fig.~\ref{fig:comrad} illustrates how the
amounts of ejecta moving inward and outward (relative to the fountain
position) change with $R$. After 150 Myr the fraction of ejecta moving
inward amounts to 20\%, 30\% and 40\% at $R=4.5$ kpc, $R=8.5$ kpc and
$R=10.5$ kpc, respectively (cf. also the thin lines in the middle
panel of Fig. \ref{fig:angmo}). Thus, this quantity increases with
$R$, in pace with the loss of angular momentum.  The fraction of the
ejecta moving outward, instead, is lower and is about the same ($\sim$
10\%) for RM and the outer fountain, but it is comparatively larger
for the inner fountain, as this one is losing less angular momentum.

\subsection{The influence of the gaseous halo}
\label{subsec:gashal}

In the previous sections we have seen how the gaseous halo influences
the circulation of the fountain gas. In this section we compare the
standard model with an analogous one where the halo is missing. This
does not mean that extraplanar gas is absent; actually, the density
of the ISM decreases along the $z$ direction following 
Eq. \ref{eq:ism} and its circular velocity depends on $z$, as illustrated
in Fig. \ref{fig:rotcu}.

As shown in the upper and middle panels of Fig. \ref{fig:nohi}, in
this model the loss of angular momentum by the ejecta is reduced
compared to RM (Fig. \ref{fig:angmo}), although a larger amount of
the fountain gas is lifted up. In fact, the absence of the coronal gas
favours the rise of the fountain, but the ejecta interacts with an
ambient medium which is also rotating (although at a lower rate than
that of the midplane), and the transfer of angular momentum is
limited. The ambient gas density at large $z$ is now higher than in
the presence of the hot halo. Radiative losses are thus facilitated,
and the ejecta does not loiter above the disk, as in SMF1, but returns
entirely back to it at later times (cf. the middle panel of
Fig. \ref{fig:nohi}). This explains why, despite a lower loss of
angular momentum, a larger fraction of ejecta moves inward compared to
RM (bottom panel). In fact, on the disk the centripetal force is
larger; as more angular momentum deficient gas falls on it, a larger
amount of ejecta move inward. Likewise, the fraction of ejecta moving
outward increases compared to RM due to the lower loss of angular
momentum that facilitates the spreading of the material outward the
injection region.

In conclusion, a smaller amount of gas (40\%) remains within
$7.5<R<8.5$ kpc compared to RM because a larger fraction of it moves
both outward and inward.

\subsection{Neglecting the neutral hydrogen}
\label{subsec:neghyd}

It is known that the neutral hydrogen distribution is rather clumpy
\citep [e.g.][]{wolf03}, instead of smooth, as we have assumed
(cf. Eq. \ref{eq:ism}). Actually, such an assumption is necessary
because an unrealistic grid resolution would be needed in order to resolve
spatially the H {\sevensize I} clouds. One can expect that the leading
shock wave of the bubble/fountain moving through a realistic ISM would
advance faster, because most of the volume is filled by the rarefied H
{\sevensize II} gas which allows higher shock velocities, as well as lower
post-shock radiative losses. To evaluate how our assumption on the H
{\sevensize I} distribution influences our results, we have run a model
similar to the reference one, but where H {\sevensize I} is absent;
this is a rather extreme assumption, but allows us to establish an upper
limit value on the heights reached by the fountain.

\begin{figure}    
\begin{center}    
\psfig{figure=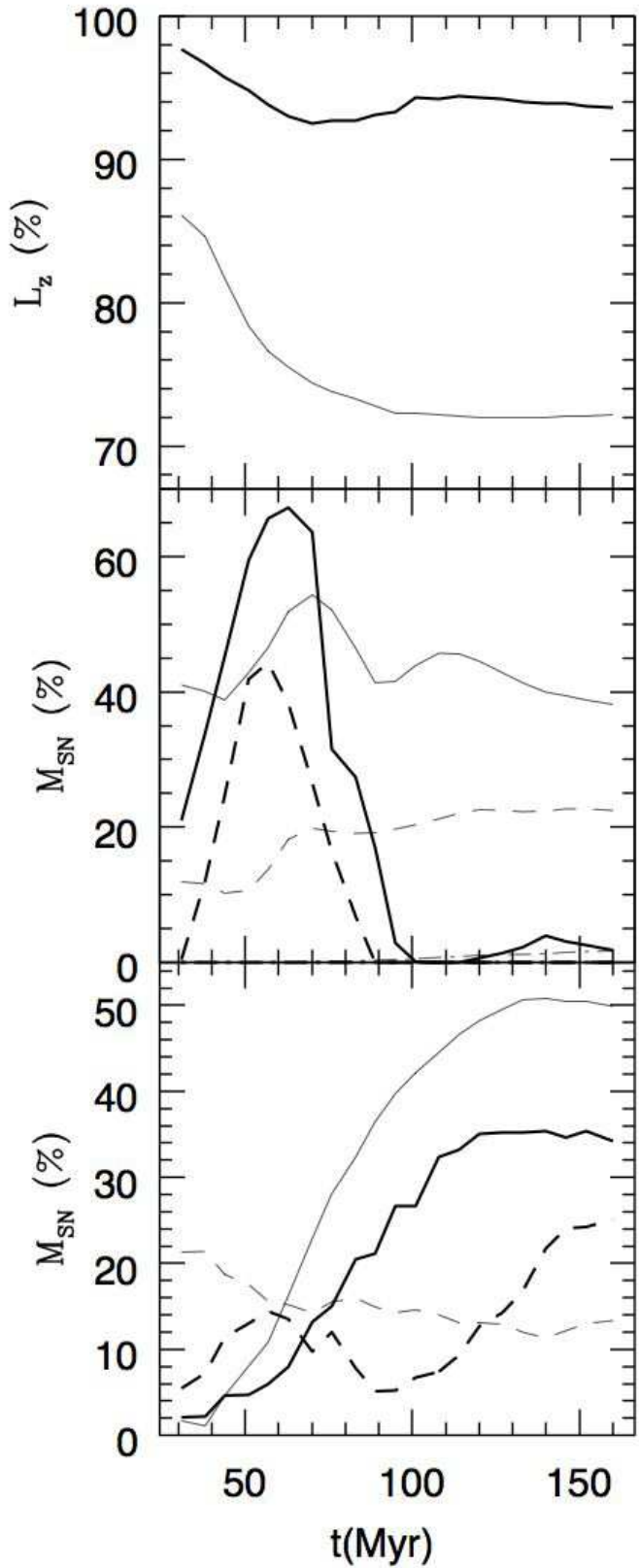,width=0.5\textwidth}    
\end{center}   
\caption{ 
Evolution of two models of fountains similar to RM, but one without
the hot gaseous halo (thick lines), and another without H {\sevensize
I} gas (thin lines). The meaning of the panels as well as that of the
line types is the same as in Fig. \ref{fig:angmo}.}
\label{fig:nohi} 
\end{figure}

Actually, after 150 Myr the amount of the ejecta above 800 pc and
above 1200 pc in the model without H {\sevensize I} is almost doubled
compared to RM (cf. the second panel of Fig. \ref{fig:nohi}).
This is not a consequence of larger heights reached by the ejecta,
because it actually attains approximately the same altitudes as in RM.
In fact, we find that the lifted ejecta does not fall back and
remains essentially at the height reached initially; at larger values
of $z$ the amount of ejecta even tends to increase with time. This behaviour
is due to the decrease of the radiative losses as a consequence of the lower
density of the ambient medium: the hot ejecta does not cool and
tends to keep moving upward by buoyancy.

As a larger fraction of ejecta interacts longer with the hot halo
compared to RM, the loss of angular momentum is also larger, as well
as the amount of ejecta moving inward (cf. the first and third panels
of Figs. \ref{fig:angmo} and \ref{fig:nohi}; this mechanism is
substantially similar to that discussed in section \ref{subsec:galdis}
with regard to the model with $R=10.4$ kpc.

In conclusion, in the absence of H {\sevensize I}, 40\% of the ejecta mass
remains above $=800$ pc, and less than 40\% is found within $7.5<R<8.5$ kpc
(characterizing a larger spreading).

\begin{figure}    
\begin{center}    
\psfig{figure=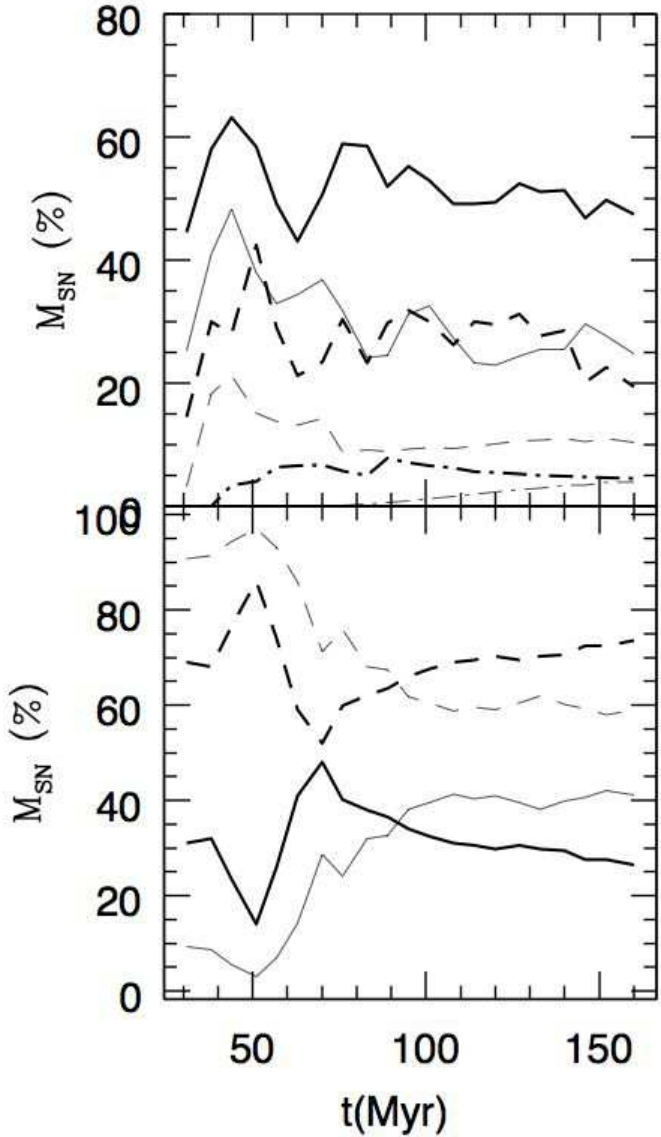,width=0.5\textwidth}    
\end{center}   
\caption{The upper panel shows the amount of the ejecta at different
  heights from the Galactic plane obtained by the RM (thin lines)
  compared with those given by a model without rotation (thick
  lines). The lower panel displays the amount of the ejecta located
  inside (dashed lines) and outside (solid lines) the cylindrical
  region above the stellar cluster powering the fountain, as well as
  the fraction located outside the cylinder (see the text for the
  details).}
\label{fig:vrot0} 
\end{figure}

\subsection{The role of rotation}
\label{subsec:rot}

Previous authors \citep[e.g.][]{deav00} have investigated numerically
the mechanism of the galactic fountains, focussing only on a
restricted region of the galactic plane; in their computational domain
all the quantities (gas and stellar distribution, as well as the
gravity) depends only on $z$. In their simulations also the Galactic
rotation is obviously absent. To evaluate the importance of the
rotation, we have run a model similar to the reference one, but where
the rotation is inhibited. In order to get a stable disk, we also
assumed that no radial dependence is present, the ISM being structured
as a homogeneous disk whose density and pressure depends only on $z$,
and the value of these variables at $z=0$, as well as the vertical
scale height, are the same found at $R=8.4$ kpc in the reference
model. The same is true for the vertical component of the
gravitational force, while its radial component is set to zero.

Figure \ref{fig:vrot0} compares results obtained in absence of rotation
with analogous results for the reference model. Both models have a maximum
resolution of 26 pc.
The upper panel shows that in the non rotating
case the ejecta reaches larger heights than in the reference model. In
fact, in the non rotating model the fountain resembles a column, and
the SNe II located at its bottom can push the gas upward more
effectively. In the rotating model, instead, the gas at larger $z$
(ejected at former times) lags systematically behind the freshly
expelled gas along the circular trajectory (because of the drag by the
gaseous halo); the fountain is thus less efficient at driving the gas
to larger heights.

The differences in the distribution of the ejecta between RM and the
non rotating model are illustrated in the lower panel of
Fig. \ref{fig:vrot0}. Contrary to the previous cases, we now do not
separate ideally the disk in an inner and an outer region; instead, we
consider a cylindrical volume extending vertically and intersecting
the disk with a circle with a diameter of 1 kpc and centred on the
cluster located at $R=8.4$ kpc. At early times the ejecta in the non
rotating model is mostly located inside the cylinder, but then
straggles outside while falling back. The ejecta in the
RM shows the same qualitative behaviour, but the mass
fraction inside the cylinder remains always higher than in the non
rotating case. The larger the heights reached by the ejecta, the
larger its diffusion: the fraction left behind in RM along a bow
shaped tail (cf. Fig. \ref{fig:colden}) is negligible and does not
play a substantial role in this analysis.

\section{Discussion and Conclusions}
\label{sec:conc}

We studied in detail the development of the Galactic fountains in
order to understand their dynamical evolution, their role in building
intermediate velocity clouds and condensations up to few kpc over
the disk, their interaction with the hot halo and their influence in
the redistribution of the freshly delivered metals over the disk. In
this paper we describe the evolution of a single fountain, i.e.  a
fountain powered by SNe II belonging to a single stellar cluster.  In
a companion paper we will present the results of simulations of
multiple fountains fuelled by SNe II of different OB associations of
various richness, occurring stochastically in space and time.
%We will study how different fountains cooperate and how they
%interact with external gas accreting onto the Galaxy.

Not all the SNe II contribute to build galactic fountains.  Only
bubbles powered by a luminosity a few times larger than the critical
luminosity $L_{\rm b}$ break out of the disk pushing the ambient gas
some kpc above the Galactic plane (cf. \ref{subsubsec:singlesnexp}).
As shown in Fig. {\ref{fig:lcr}, this critical luminosity
decreases exponentially with the Galactocentric distance. In the next
paper we will show how, combining
this behaviour with the cumulative probability for the occurrence of
$N_{\rm SN}$ clustered supernovae \citep{hili05}, it turns out that
only $\sim 60$\% of all SNe II occurring on the disk can give rise to
Galactic fountains. This fraction reduces almost linearly with radius;
for instance, it becomes $\sim 40$\% within $R=4$ kpc.  It is likely
that, in the case of multiple fountains, also weaker superbubbles may
reach larger heights above the plane finding their way through breaks
and tunnels carved by previous, adjacent and more powerful
superbubbles. A discussion on this point is postponed in the companion
paper.

In any case, a non-negligible fraction of superbubbles seems to remain
contained by the thickened Galactic disk and do not participate to
a large scale gas circulation. The fate of such superbubbles has been
envisaged by \citet{teno96} who suggests that the superbubble stops
its expansion stalling at an height $z\sim 1$ kpc. The hot,
low-density, metal-rich gas of its interior waits for the onset of
radiative cooling. Small overdense inhomogeneities cool faster
increasing their density and forming droplets which start to fall
toward the disk. This rain of small ($\sim 0.1$ pc), metal-rich
cloudets occurs over a kpc-scale volume of the galaxy. These cloudets
will be successively photo-evaporated by the stellar radiation,
eventually mixing with the ISM. Although some objection may be raised
about the details of this scheme \citep{recchi01}, there is no doubt
that a stalled bubble must deflate as a consequence of the
depressurization of its interior due to radiative losses. It is also
possible that the turbulences of the local medium distort and destroy
the superbubble, anticipating its disappearance.
In conclusion, weak bubbles do not affect much the general
distribution of the ambient medium. Their size can be rather small, of
the order of 100 pc (cf. the left panel of Fig. \ref{fig:rir2}), and
their metal content remains strongly localized.

On the contrary, powerful superbubbles give rise to Galactic
fountains, and are expected to drive a gas circulation on large scale
\citep{shafi76, breg80}. Simple ballistic simulations described in the
Appendix \citep[see also][]{frbi06}, where the clouds are treated as
bullets ejected into the halo from the Galactic midplane, actually
show that the clouds fall back on the disk at distances
that can be as large as several kpc from the starting point, depending on the
modulus and direction of the initial velocity (cf. the Appendix).  Our
hydrodynamical simulations, however, show that the majority of the gas
lifted up by the fountains falls back on the disk remaining within a
radial distance $\Delta R=0.5$ kpc from the place where the fountain
originated.  Our reference model, a fountain powered by 100 SNe II
occurring at $R=8.5$ kpc, may eject material up to $\sim 2$ kpc which
than collapses back mostly in form of dense, cold clouds and
filaments. Although the simple ballistic model indicates that these
cloud should move mostly outward (cf.  the Appendix), our simulations
show instead that they have the tendency to drift toward the Galactic
centre as they approach the disk. This effect is due to the
interaction between the fountain and the gaseous halo above the
disk. This halo is at rest and exerts a drag force on the rotating
fountain gas which looses part of its angular momentum and is pulled
inward by the gravitational force.
Nearly 60\% of the metals delivered by the SNe II remains in the
radial range $8<R<9$ kpc, while the rest is dispersed on a larger
range, mostly inward rather than outward (c.f. Fig. \ref{fig:angmo}).
We stress that even in absence of the gaseous halo the fraction of
metals moving inward still prevails over that moving outward, although
to a smaller extent (c.f.  Fig. \ref{fig:nohi}). In fact, the
exponential tail of the vertical distribution of the disk gas is dense
enough to exert a non-negligible drag on the fountain.

We point out that our results are different from those of \citet{both}
who obtain much longer trajectories for the clouds pushed by the
fountains.  As an example, these authors find that the gas ejected at
$R=8,5$ kpc next passes through the disk at $R=25$ kpc (cf. their
Fig. 5). \citet{both} adopt an SPH code and clouds are modelled using
`sticky particles' that move ballistically through the ambient
gas. However, also in the framework of pure ballistic trajectories
(which is the most favourable to obtain large radial displacements) it
is difficult to reconcile the results of \citet{both} with the
shorter trajectories illustrated in our Appendix.

We also stress that our results about the radial displacement of the
metals are in agreement with recent chemical models of the Milky Way
aimed to explain the chemical gradient along the Galactic disk
\citep{cemafr07}. In these models the disk is approximated by several
independent rings, 2 kpc wide, without exchange of matter between
them. Because such models are successful in reproducing the observed
gradients, one can infer that most of the metals do not move radially
more than 1 kpc from the place where they are created, as found by our
simulations.

We finally make some remarks about the interaction between the
fountains and the hot gas of the halo. The SN II feedback, infact, is
an important element to understand the energy budget of the halo
\citep[e.g.][]{wa05}. Unfortunately, at this stage, a definite
conclusion can not be drawn by our models of single fountains. The
amount of thermal energy of the gas of the halo in our Galactic model
is $E_{\rm th}\sim 10^{55}$ erg. Although this gas is set initially at
rest, some angular momentum is transferred to it by the rotating ISM
of the disk through numerical viscosity. While the amount of this
angular momentum is small (and the induced rotation is highly
subsonic), it introduces some noise at the vertical edges of our
computational box, and the thermal energy of the gas reduces of an
amount $\Delta E_{\rm th}=3\times 10^{53}$, about 3\% of the
total. This energy must be compared with $E_{\rm fnt}=5\times 10^{52}$
erg, which is the energy injected in the numerical grid by the SNe
II. This energy is ten times lower than the perturbation generated at
the boundaries, and any evaluation of the energy transfer from the
fountain to the halo is prevented. This subject will be toroughly
investigated in the companion paper on the multiple fountains,
when the energy injected by the SNe is $\sim$ $10^{54}$ erg and we
will be able to give a reasonably accurate evaluation of the fraction
of energy transferred to the halo gas.

\section*{Acknowledgments}

We are indebted to an (anonymous) referee and to Alex Raga for their
extremly profitable comments and insightful suggestions that have
greatly helped to improve this manuscript. E.M.G.D.P. also
acknowledges the partial support from grants of the Brazilian
Agencies FAPESP and CNPq.

%\begin{thebibliography}{99}
%\end{thebibliography}

\bibliographystyle{mn2e} \bibliography{fontane_ref}

\appendix
\section{The ballistic case}

To compare the hydrodynamical description of a single fountain given
in this paper with the pure kinematics picture given by \citet{breg80},
we have built a model to follow the orbits of gas clouds that are
ejected from the disk and move ballistically within the same potential
well adopted in our hydrodynamical models (see sect. \ref{subsec:galmod}). 

In this simple model we integrate the trajectory of only a single
cloud which is modelled as a particle initially moving around
circular orbits in the plane $z=0$ and receiving a kick velocity
aloft.  The integration is performed in cartesian coordinates using
the fourth-order Runge-Kutta method, and the trajectory is followed
until the cloud falls back to the disk.

In all the cases presented in Fig. \ref{fig:bal} the cloud
starts at $R=8.5$ kpc with a kick velocity $v_{\rm ej}=10^2$ km
s$^{-1}$. The cases differ only in the direction of $v_{\rm ej}$. In the
bottom panel of Fig. \ref{fig:bal} three trajectories in the $z-R$
plane are shown for which $v_{\rm ej}$ is perpendicular to the
circular velocity $v_{\rm c}$, but have three different inclinations
with respect to the Galactocentric distance $R$. The three
trajectories in the top panel refer to cases
where $v_{\rm ej}$ is orthogonal to $R$ but with different slope angles to
$v_{\rm ej}$. The case of orthogonal ejection velocity to the Galactic
plane is represented in both panels for an easier comparison.
 
\begin{figure*}    
\begin{center}    
\psfig{figure=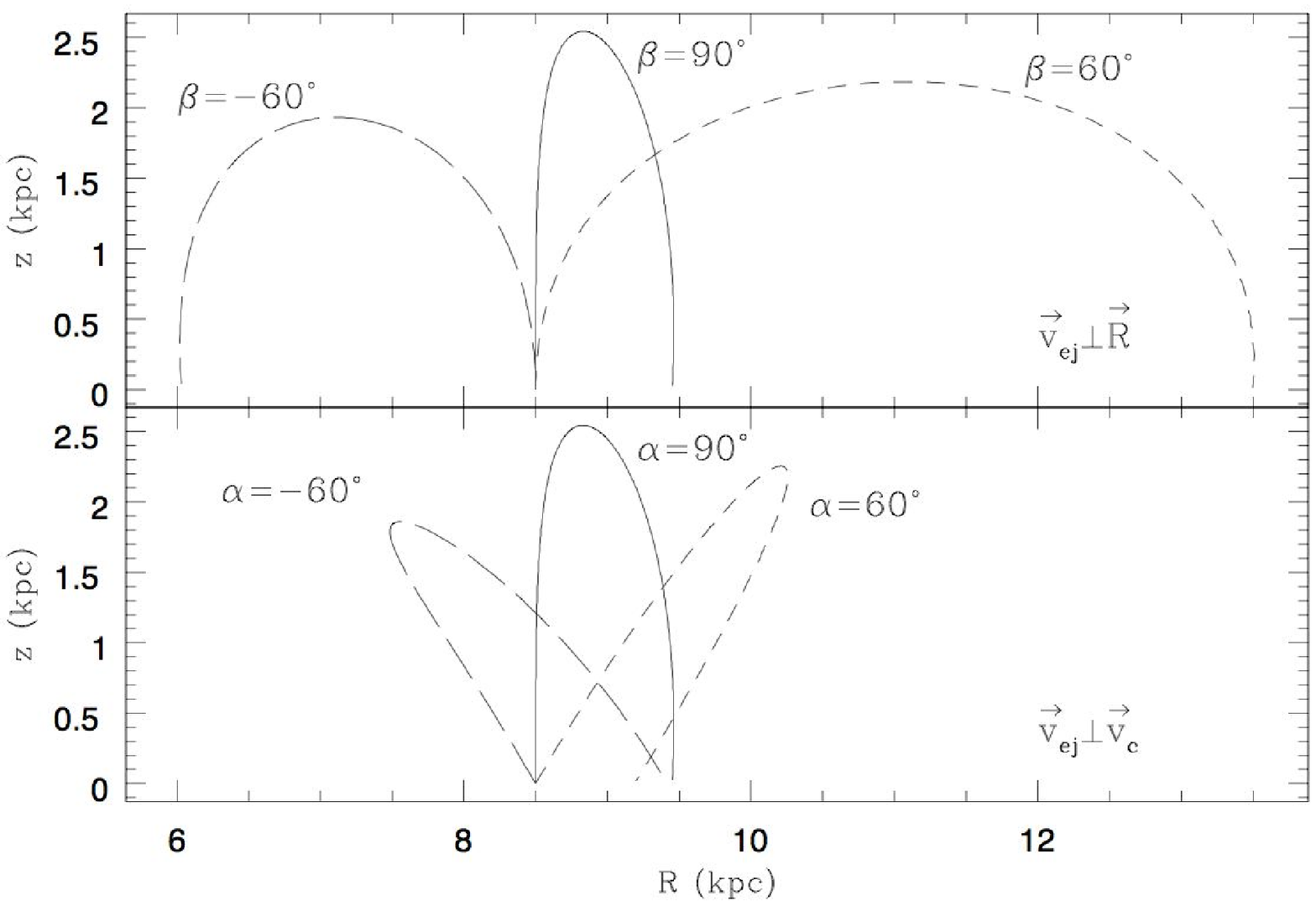,width=1.0\textwidth}    
\end{center}   
\caption{ 
Trajectories of a single cloud in the $z-R$ plane, all starting at 
$R=8.5$ kpc with a kick velocity $v_{\rm ej}=10^2$ km s$^{-1}$. See
the text for the details.
  }
\label{fig:bal} 
\end{figure*}

Figure \ref{fig:bal} emphasizes the importance of the initial
conditions.  In the cases illustrated in the bottom panel, despite the
very different orbits the cloud falls back on the plane essentially at
the same Galactcentric distance, not far from the starting point. In
the examples shown in the top panel, instead, the cloud can ''land''
much closer to, or quite further away, the Galactic centre. In these cases
one can expect consequences in the kinematics of the thick disk, as
well as in the stellar abundance gradients along the disk.

\label{lastpage}
\end{document}